# Quantum thermometric sensing: Local vs. Remote approaches

Seyed Mohammad Hosseiny ,[1, *] Abolfazl Pourhashemi Khabisi,[1] Jamileh Seyed-Yazdi
,[1, †] Milad Norouzi ,[1] Somayyeh Ghorbani,[2, ‡] Asad Ali ,[3] and Saif Al-Kuwari [3]

[1]*Physics Department, Faculty of Science, Vali-e-Asr University of Rafsanjan, Rafsanjan, Iran*
[2]*Physics Department, Faculty of Sciences, Urmia University, P.B. 165, Urmia, Iran*
[3]*Qatar Centre for Quantum Computing, College of Science and Engineering, Hamad Bin Khalifa University, Doha, Qatar*
(Dated: October 17, 2025)

**Abstract** Quantum thermometry leveraging quantum sensors is investigated with an emphasis on fundamental precision bounds derived from quantum estimation theory. The proposed sensing platform consists of two dissimilar qubits coupled via capacitor, which induce quantum oscillations in the presence of a thermal environment. Thermal equilibrium states are modeled using the Gibbs distribution. The precision limits are assessed through the Quantum Fisher Information (QFI) and the Hilbert-Schmidt Speed (HSS), serving as stringent criteria for sensor sensitivity. Systematic analysis of the dependence of QFI and HSS on tunable parameters—such as qubit energies and coupling strengths—provides optimization pathways for maximizing temperature sensitivity. Furthermore, we explore two distinct quantum thermometry paradigms: (i) local temperature estimation directly performed by Alice, who possesses the quantum sensor interfacing with the thermal bath, and (ii) remote temperature estimation conducted by Bob, facilitated via quantum teleportation. In the latter scenario, temperature information encoded in the qubit state is transmitted through a single-qubit quantum thermal teleportation protocol. Our findings indicate that direct measurement yields superior sensitivity compared to remote estimation, primarily due to the inherent advantage of direct sensor-environment interaction. The analysis reveals that increasing Josephson energies diminishes sensor sensitivity, whereas augmenting the mutual coupling strength between the qubits enhances it.

## I. INTRODUCTION

Temperature measurement has undergone a paradigm shift: from traditional macroscopic techniques—such as the expansion of mercury columns or the resistance variation of thermistors—to direct quantum-state readout of individual quantum systems, including atoms, vacancies, or superconducting qubits. Over the past decade, *quantum sensors*—such as nitrogen-vacancy (NV) centers in diamond [1–4], cold-atom impurities [5–8], polarons in Bose–Einstein condensates [9–11], and engineered on-chip spin chains [12–14]—have evolved into primary thermometric tools, with their fundamental precision bounded solely by the quantum Cramér–Rao bound [15, 16]. These devices can resolve millikelvin temperature variations within millisecond timescales, spanning from picokelvin regimes in dilution refrigerators to physiological temperatures at approximately 37 °C, often occupying sub-cellular volumes [17].

The central idea is simple yet radical [18–24]: a well-characterised quantum probe (two-level systems such as qubits) is brought into thermal contact with the sample; its equilibrium or dynamical observables encode the temperature. Exploiting entanglement or nonequilibrium [25] drive can further suppress noise, allowing *sub-nanokelvin* sensitivities in ultracold gases [10] and *sub-micron spatial resolution* in dissipative quantum devices [26, 27]. Recent theoretical work has established general strategies for optimising the probe Hamiltonian, the interaction time, and even the measurement sequence, while experimental milestones include NV-based intracellular thermometry [28, 29] and on-chip thermal imaging of superconducting circuits [30, 31].

In this article we study the rapidly maturing field of *quantum thermometry by quantum sensors*. We begin by deriving the ultimate precision bounds set by quantum estimation theory [18, 32–37], then survey state-of-the-art platforms—two dissimilar-charge qubits—and conclude with emerging applications that range from cryogenic quantum processors [38, 39] to early cancer diagnostics [40].

This work explores the optimization of a quantum sensor in order to quantum thermometry [18, 21, 22, 41]. The sensor consists of two charge-dissimilar qubits coupled electrostatically via an on-chip capacitor and interacting with a thermal bath at a certain temperature $T$. The equilibrium state is considered by a Gibbs thermal state of the overall Hamiltonian. The quantum sensor operates as a probe to estimate temperature for quantum thermometry. We use the quantum Fisher information (QFI) [32, 42] and Hilbert-Schmidt speed (HSS) [43, 44] to assess precision limits, specifically focusing on temperature estimation to improve the sensor's quantum sensitivity. Notably, the QFI and HSS quantify the precision of a quantum thermometry scheme independent of specific measurement protocols or experimental setups [18]. To this end, we analyze the impact of key parameters such as Josephson energies of qubits and mutual coupling energy between qubits on both QFI and HSS in quantum thermometry. Our goal is to identify optimal working regimes where the quantum sensor exhibits enhanced sensitivity to temperature variations. We present numerical results illustrating the behavior of QFI and HSS as a function of these parameters, offering insights into the design and optimization of quantum thermometers based on capacitively coupled dissimilar qubits. Furthermore, we discuss the improvement of the performance characteristics of this quantum sensor.

We also compare the sensitivity of a quantum sensor to that of quantum remote sensing in estimating temperature during quantum teleportation [45–52], where temperature information is transmitted via a quantum state between distant sender (Alice) and receiver (Bob). In essence, we examine two quantum thermometry scenarios: one where Alice, possessing the quantum sensor, estimates the temperature directly; and another where Alice sends the qubit state via the sensor's quantum state to Bob, enabling him to remotely estimate the temperature at the teleportation destination. One of the very important and practical results mentioned in this article is that the sensitivity of temperature estimation in the first scenario, where Alice directly estimates the temperature using the quantum sensor, is higher than in the quantum teleportation destination, where Bob estimates it remotely. To delve deeper, this advantage in sensitivity arises from the direct interaction between the quantum sensor and the environment in Alice's scenario. This direct coupling allows for a more precise measurement of the temperature. In contrast, the teleportation process, while faithfully transferring the quantum state, introduces inherent noise and imperfections that degrade the sensitivity of Bob's temperature estimation. It reveals a key trade-off between direct sensing and remote state transfer in quantum metrology. Moreover, we demonstrate that tuning of our sensor parameters is crucial for thermometry, such that increasing Josephson energies of qubits suppresses the sensitivity of the quantum sensor, while increasing mutual coupling energy between qubits enhances it. The applications of this article cover a wide range, including: ultra-precise cryogenic thermometry [53], in vivo nano-thermometry [28], on-chip hot-spot mapping [54], distributed environmental monitoring [53], and point-of-use thermal validation [55].

The paper is structured into four main sections: after the introduction, Section II elucidates the fundamentals of quantum teleportation. Section III delves into the intricacies of two dissimilar coupled qubits that demonstrate quantum oscillations employing pulse techniques. Finally, Section IV summarizes and scrutinizes the core findings of the study.

## II. PRELIMINARIES

### A. Quantum teleportation

Quantum teleportation transfers quantum information about an unknown state from one location (Alice) to another distant location (Bob) [45–52]. This process relies on a channel connecting the sender and receiver. This channel is composed of two ingredients: a classical communication channel and a quantum communication channel, which is a pre-shared entangled quantum state between Alice and Bob. The classical channel is used to transmit the results of Alice's measurements on her portion of the entangled state and the unknown quantum state. Alice performs a Bell measurement on her qubit and the qubit containing the unknown quantum state, projecting them onto one of four Bell states. The result of this measurement is then communicated to Bob via a classical channel. Conditioned on Alice's measurement outcome, Bob then uses this information to perform a specific quantum operation on his portion of the entangled state (such as applying a specific quantum gate to his qubit), thereby reconstructing the original unknown quantum state. It's crucial to understand that quantum teleportation doesn't involve transferring matter or energy; instead, it's the quantum information itself that's teleported. Furthermore, the no-cloning theorem remains intact, as Alice's measurement destroys the original quantum state at her location.

In the standard teleportation protocol, Bell measurements and Pauli rotations are used as local quantum operations. Bowen and Bose [56] demonstrated that the standard teleportation protocol, using mixed states as a resource $\rho_{ch}$, is equivalent to a generalized depolarizing channel $\Lambda(\rho_{ch})$ with the input state $\rho_{in} = |\psi_{in}\rangle\langle\psi_{in}|$. The unknown initial state of single-qubit quantum teleportation for any arbitrary pure single-qubit state can be considered as $|\psi_{in}\rangle = \cos\left(\frac{\theta}{2}\right)|0\rangle + e^{i\phi}\sin\left(\frac{\theta}{2}\right)|1\rangle$, where $\theta$ and $\phi$ are the amplitude and phase of the initial state of the quantum teleportation, respectively. Thus, the teleportation output state $\rho_{out}$ for an arbitrary single-qubit input state $\rho_{in}$ is obtained as follows [56]:

$$\rho_{out} = \Lambda(\rho_{ch})\rho_{in} = \sum_{i=0}^{3} \text{Tr}[\mathcal{B}_i\rho_{ch}]\sigma_i\rho_{in}\sigma_i, \tag{1}$$

where

$$\mathcal{B}_i = \left(\sigma_0 \otimes \sigma_i\right)\mathcal{B}_0\left(\sigma_0 \otimes \sigma_i\right), \quad i = 1, 2, 3, \tag{2}$$

represents the Bell state associated with the Pauli matrix $\sigma_0 = \mathbb{I}, \sigma_1 = \sigma_x, \sigma_2 = \sigma_y, \sigma_3 = \sigma_z$ and $\mathbb{I}$ denotes the identity matrix. In the standard base $\{|0\rangle, |1\rangle\}$, for any two arbitrary qubits we have $\mathcal{B}_0 = \frac{1}{2}\left(|00\rangle + |11\rangle\right)\left(\langle00| + \langle11|\right)$.

The fidelity criterion assesses the similarity between input and output (teleported) states. It is a crucial metric in quantum teleportation, quantifying how accurately the original quantum state is transferred to the receiver. High fidelity indicates a successful teleportation, where the output state closely resembles the input state. Imperfections in the teleportation protocol, such as noise or imperfect entanglement, can reduce the fidelity. Thus, the fidelity criterion $f(\rho_{in}, \rho_{out})$ can be defined as [57, 58]:

$$f(\rho_{in}, \rho_{out}) = \left( \mathrm{Tr}\left[ \sqrt{ \sqrt{\rho_{in}} \rho_{out} \sqrt{\rho_{in}} } \right] \right)^2.$$

(3)

where the fidelity limit is $0 \leq f(\rho_{in}, \rho_{out}) \leq 1$. For $f = 1$, the optimal fidelity required to achieve successful teleportation can be attained. The threshold of the maximum classical fidelity occurs at $f = 2/3$; afterward, we go into the quantum fidelity.

## B. Quantifying the metrological precision of a quantum sensor based on the quantum estimation theory

### 1. Quantum Fisher information

The QFI quantifies the precision of a thermometry scheme independent of specific measurement protocols or experimental setups [18]. It provides a fundamental limit on the sensitivity achievable in estimating temperature. In essence, a higher QFI indicates a greater potential for precise temperature measurements.

This section provides a concise overview of single-parameter quantum estimation theory, introducing key concepts and tools used in this study. In quantum metrology, the QFI quantifies the maximum achievable precision for estimating an unknown parameter $\vartheta$. The Cramér–Rao (CR) bound determines the ultimate precision limit for any quantum state $\rho_\vartheta$, where the variance of any estimator $\hat{\vartheta}$ is lower bounded by the reciprocal of the QFI [32, 33, 59, 60]:

$$\mathrm{Var}(\vartheta) \geq \frac{1}{m \mathcal{F}(\vartheta)},$$

(4)

where $\mathrm{Var}(\vartheta)$ represents the variance, $m$ denotes the number of measurements repeated, such that $m = 1$ is a single-shot scenario, and $\mathcal{F}(\vartheta)$ denotes the QFI corresponding to the quantum state $\rho_\vartheta$ and its parameterization process. The QFI is maximized over all possible positive-operator-valued measures (POVMs) [33] to determine the ultimate bound on the precision with respect to the parameter $\vartheta$. The QFI for a mixed state can be defined by:

$$\mathcal{F}(\vartheta) = \mathrm{Tr}\left[\rho_\vartheta L_\vartheta^2\right],$$

(5)

in which $L_\vartheta$ represents the symmetric logarithmic derivative (SLD) which is given by

$$\frac{\partial \rho_\vartheta}{\partial \vartheta} = \frac{L_\vartheta \rho_\vartheta + \rho_\vartheta L_\vartheta}{2},$$

(6)

Considering $L_\vartheta$ in the eigenbasis of $\rho_\vartheta$, one can rewrite the QFI as follows:

$$\mathcal{F}(\vartheta) = 2 \sum_{n,m} \frac{|\langle \psi_n | \partial_\vartheta \rho_\vartheta | \psi_m \rangle |^2}{\lambda_n + \lambda_m}.$$

(7)

here, $\lambda_{n,m}$ and $|\psi_{n,m}\rangle$ represent the eigenvalues and eigenvectors of the density matrix $\rho_\vartheta$, respectively. Further, $\partial_\vartheta = \partial/\partial\vartheta$ denotes the partial derivative corresponding to $\vartheta$.

The SLD is defined using the density matrix $\rho_\vartheta$, whose eigenvalues and eigenvectors may depend on the parameter $\vartheta$. Applying the Leibniz rule, we can decompose the QFI contribution into two parts.

$$\partial_\vartheta \rho_\vartheta = \sum_n \left( (\partial_\vartheta \rho_\vartheta) |\psi_n\rangle \langle\psi_n| + \rho_\vartheta |\partial_\vartheta \psi_n\rangle \langle\psi_n| + \rho_\vartheta |\psi_n\rangle \langle\partial_\vartheta \psi_n| \right),$$

(8)

where,

$$|\partial_\vartheta \psi_n\rangle = \sum_k \partial_\vartheta \psi_{nk} |k\rangle,$$

(9)

where $\{|k\rangle\}$ is an arbitrary basis. Assuming $\langle \psi_n | \psi_m \rangle = \delta_{nm}$, one can write

$$\partial_\vartheta \langle \psi_n | \psi_m \rangle \equiv \langle \partial_\vartheta \psi_n | \psi_m \rangle + \langle \psi_n | \partial_\vartheta \psi_m \rangle = 0$$

(10)

and, we have

$$\text{Re}\langle\partial_\vartheta\psi_n\mid\psi_m\rangle=0,\quad\langle\partial_\vartheta\psi_n\mid\psi_m\rangle=-\langle\psi_n\mid\partial_\vartheta\psi_m\rangle=0. \tag{11}$$

Utilizing Eq. 8 and the above relations, one can obtain

$$L_\vartheta=\sum_p\frac{\partial_\lambda\lambda_p}{\lambda_p}\mid\psi_p\rangle\langle\psi_p\mid+2\sum_{n\neq m}\frac{\lambda_n-\lambda_m}{\lambda_n+\lambda_m}\langle\psi_m\mid\partial_\vartheta\psi_n\rangle\mid\psi_m\rangle\langle\psi_n\mid, \tag{12}$$

such that the QFI can be written as

$$\mathcal{F}(\vartheta)=\sum_p\frac{\left(\partial_\vartheta\lambda_p\right)^2}{\lambda_p}+2\sum_{n\neq m}\frac{(\lambda_n-\lambda_m)^2}{\lambda_n+\lambda_m}\mid\langle\psi_m\mid\partial_\vartheta\psi_n\rangle\mid^2. \tag{13}$$

The first term represents the classical contribution to the QFI, and the second represents the purely quantum contribution.

### 2. Quantum Hilbert-Schmidt speed

The quantum statistical speed is a fundamental concept in quantum information theory, playing a crucial role in diverse areas such as quantum metrology and quantum state discrimination. We here offer a computationally tractable approach to explore the dynamic properties of quantum systems. The HSS, derived from the Wigner-Yanase metric, provides a sensitive measure of distinguishability between infinitesimally close quantum states, reflecting the system's capacity to change. We will outline the mathematical formulation of the HSS within the density matrix formalism, establishing its connection to the underlying Hamiltonian and its eigenstates. We begin by reviewing the general framework for defining the HSS, a special case of quantum statistical speed, within a family of distance measures [43]:

$$[d_\alpha(p,q)]^\alpha=\frac{1}{2}\sum_x\mid(p_x)^{\frac{1}{\alpha}}-(q_x)^{\frac{1}{\alpha}}\mid^\alpha, \tag{14}$$

where $p=\{p_x\}_x$ and $q=\{q_x\}_x$ are probability distributions. The aforementioned relation is parameterized by $\alpha\geq1$. taken from a single-parameter family $p_x(\vartheta_0)$ with parameter $\vartheta$. Therefore, the classical statistical speed can be defined as

$$s_\alpha[p(\vartheta_0)]=\frac{d}{d\vartheta}d_\alpha\big(p(\vartheta_0+\vartheta),p(\vartheta_0)\big), \tag{15}$$

Assuming two quantum states $\rho$ and $\sigma$, one can extend the classical notions to the quantum situation by taking as the measurement $p_x=\text{Tr}[E_x\rho]$ and $q_x=\text{Tr}[E_x\sigma]$ probabilities related to the POVMs determined by the set of $\{E_x\geq0\}$ fulfilling $\sum_n p_x=\mathbb{I}$, in which $\mathbb{I}$ represents the identity operator. Thus, the quantum distance that maximizes the classical distance over all POVMs can be given by:

$$D_\alpha(\rho,\sigma)=\max_{E_x}d_\alpha(\rho,\sigma), \tag{16}$$

In the quantum setting, maximizing the classical distance in Eq. (14) over all POVMs, as informed by Eq. (15), gives the quantum statistical speed, or Hilbert–Schmidt distance as:

$$S_\alpha[\rho_\vartheta]=\max_{E_x}s_\alpha[p(\phi)]=\left(\frac{1}{2}\text{Tr}\left[\frac{d\rho_\vartheta}{d\vartheta}\right]^\alpha\right)^{1/\alpha}. \tag{17}$$

For case $\alpha=2$, the quantum statistical speed is reduced by the HSS as

$$\text{HSS}(\rho_\vartheta)=\left(\frac{1}{2}\text{Tr}\left[\frac{d\rho_\vartheta}{d\vartheta}\right]^2\right)^{1/2}. \tag{18}$$

where it does not need to diagonalize the $d\rho_\vartheta/d\vartheta$. It only needs to evaluate it at a specific $\vartheta$.

In the continuation, after introducing our quantum sensor, we investigate the single-parameter estimation for temperature as a noise source in quantum thermometry by QFI and HSS tools. To this end, we consider a quantum sensor interacting with a thermal reservoir at an unknown temperature $T$. We employ the QFI and HSS as figures of merit to assess the precision of temperature estimation.

## III. QUANTUM SENSOR

We present a quantum sensor for temperature measurement comprising two charge-dissimilar qubits electrostatically coupled via an on-chip capacitor $C_M$ (see Fig. 1). The first qubit, with a superconducting quantum interference device (SQUID) geometry, enables control of the Josephson coupling to its reservoir, leading to quantum oscillations. This asymmetry between qubits structure introduces a controllable parameter, the magnetic flux through the SQUID, which allows for tuning the energy levels of the qubit and manipulating its quantum state. The presence of quantum oscillations is a direct consequence of the superposition principle in quantum mechanics. When the qubit is prepared in a superposition of its ground and excited states, the probability of finding the qubit in either state oscillates in time. The frequency of these oscillations, known as the Rabi frequency, is determined by the energy difference between the states and the strength of the driving field. Both qubits share a common pulse gate but have separate dc gates, probes, and reservoirs. The pulse gate couples equally to each box. We consider our sensor to be in thermal equilibrium with a thermal bath at a certain temperature $T$. Considering $\hbar = 1$, one can determine the system's Hamiltonian in the two-qubit basis $\{|00\rangle, |10\rangle, |01\rangle, |11\rangle\}$ as [61]:

$$H = \begin{bmatrix} E_{00} & -\frac{1}{2}E_{J1} & -\frac{1}{2}E_{J2} & 0 \\ -\frac{1}{2}E_{J1} & E_{10} & 0 & -\frac{1}{2}E_{J2} \\ -\frac{1}{2}E_{J2} & 0 & E_{01} & -\frac{1}{2}E_{J1} \\ 0 & -\frac{1}{2}E_{J2} & -\frac{1}{2}E_{J1} & E_{11} \end{bmatrix}, \tag{19}$$

where $E_{Ji}$ is the Josephson energy of the $i$-th superconducting qubit. Moreover, $E_{00}, E_{10}, E_{01}$, and $E_{11}$ represent the electrostatic energies, such that the total system electrostatic energies are defined by $E_{n1n2} = E_{c1}(n_{g1} - n_1)^2 + E_{c2}(n_{g2} - n_2)^2 + E_m(n_{g1} - n_1)(n_{g2} - n_2)$, where $E_{c1}$ and $E_{c2}$ denote the charging energies of the first and second qubits, respectively. Besides, $E_m$ denotes the mutual coupling energy between both qubits.Further, $(n_1, n_2{=}0,1)$ denote the number of excess Cooper pairs in the boxes. Additionally, $n_{g1}, n_{g2}$ represent the normalized charges induced from the related qubits by $dc$ and the electrodes of the pulse gate [61]. When the system is non-adiabatically driven to $n_{g1}{=}0.5$ or $n_{g2}{=}0.5$, it oscillates as a single qubit between degenerate states at a frequency $\omega_{1,2} = E_{J1,2}/\hbar$ [61], consistent with noise insensitivity. In this regime, we have $n_{g1} = n_{g2}{=}0.5$, $E_{00} = E_{11}$ and $E_{01} = E_{10}$. Quantum oscillations occur when $E_{J1} > E_{J2}$ or $E_{J1} < E_{J2}$.

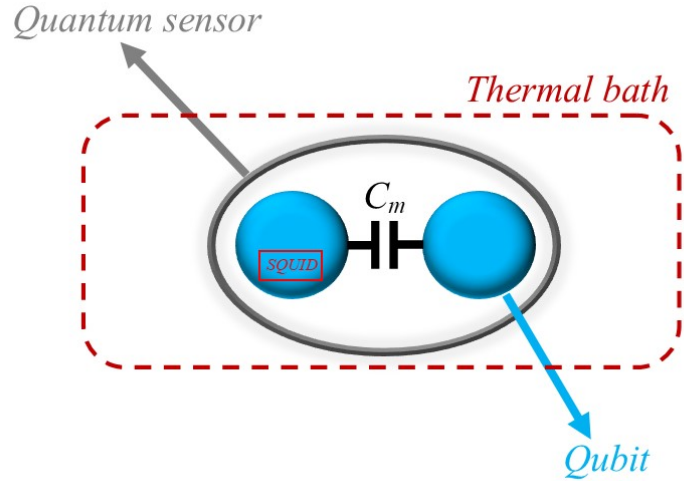

FIG. 1. A schematic of our quantum sensing protocol. The quantum sensor consists of two dissimilar qubits electrostatically coupled via a capacitor $C_M$. The first qubit has an SQUID, while the second qubit does not, leading to quantum oscillations. Our sensor is to be in thermal equilibrium with a thermal bath at a certain temperature $T$.

Thus, we can easily compute the eigenvalues $\epsilon_n$ and the eigenvectors $|\psi_n\rangle$ corresponding to the Hamiltonian (relation 19) as follows:

$$\epsilon_{1,2} = \frac{1}{4}(E_{c1} + E_{c2} \mp R_1), \quad \epsilon_{3,4} = \frac{1}{4}(E_{c1} + E_{c2} \mp R_2). \tag{20}$$

and

$$
\begin{aligned}
|\psi_{1,2}\rangle &= -|00\rangle - \left(\frac{E_m + R_1}{2(E_{J1} - E_{J2})}\right)|01\rangle + \left(\frac{E_m \pm R_1}{2(E_{J1} - E_{J2})}\right)|10\rangle + |11\rangle, \\
|\psi_{3,4}\rangle &= |00\rangle + \left(\frac{E_m + R_2}{2(E_{J1} + E_{J2})}\right)|01\rangle + \left(\frac{E_m \pm R_2}{2(E_{J1} + E_{J2})}\right)|10\rangle + |11\rangle,
\end{aligned}
\tag{21}
$$

in which $R_1 = \sqrt{4(E_{J1} - E_{J2})^2 + E_m^2}$, $R_2 = \sqrt{4(E_{J1} + E_{J2})^2 + E_m^2}$. Notably, we use $|0\rangle = \begin{pmatrix} 1 \\ 0 \end{pmatrix}$ and $|1\rangle = \begin{pmatrix} 0 \\ 1 \end{pmatrix}$ as basis states, where $|ij\rangle$ denotes $|i\rangle \otimes |j\rangle$ and $i, j = 0, 1$.

This work explores using a proposed quantum sensor to measure the thermal bath temperature $T$ (see Fig. 1). Unlike other quantum sensing schemes, ours relies on thermalization in a standard thermodynamic scenario, where the sensor equilibrates with the bath at temperature $T$, without external drives or strong nonlinear coupling. Our work explores the quantum sensor near thermal equilibrium, where this condition is more representative of actual experiments.

Assuming the system is coupled to a low-temperature bath, the quantum sensor reaches thermal equilibrium, and its state is well-described by the Gibbs thermal state:

$$
\rho_{ch}(T) = \frac{1}{z}\left(\exp(-\beta H)\right) = \frac{1}{z}\sum_n \exp(-\beta\epsilon_n)|\psi_n\rangle\langle\psi_n|.
\tag{22}
$$

where $\epsilon_n$ and $|\psi_n\rangle$ represent the eigenvalues and eigenvectors of the Hamiltonian, respectively. Moreover, $z = \mathrm{Tr}[\exp(-\beta H)]$ denotes the partition function, and $\beta = 1/(k_B T)$ (in which the $k_B$ is the Boltzmann constant such that for simplicity we consider $k_B = 1$ and $T$ is temperature). We use Eq. 22 as a resource for single-qubit quantum thermal teleportation.

The QFI and HSS are used to analyze the sensor's ability to estimate temperature $T$. We numerically compute them using the eigenvalues and eigenvectors of $\rho_{ch}(T)$ to quantify the precision of temperature estimation. Notably, we extract the straightforward expression of the thermal density matrix $\rho_{ch}(T)$ in Appendix A.

## IV. QUANTUM THERMOMETRY

At first, we consider the quantum thermometry scenario in which Alice sends the temperature information encoded on the self qubit state via the sensor's quantum state to Bob, enabling him to remotely estimate the temperature at the single-qubit thermal teleportation destination. We now calculate the output state of single-qubit quantum thermal teleportation using the channel state (Eq. A-1) and Eq. (1) as

$$
\rho_{out}(T) = \begin{pmatrix} \frac{F_1 - (B_1 + B_2)\cos(\theta)}{2F_1} & -\frac{\sin(\theta)\left(F_2\cos(\phi) + i(B_1 - B_2)\sin(\phi)\right)}{2F_1} \\ -\frac{\sin(\theta)\left(F_2\cos(\phi) - i(B_1 - B_2)\sin(\phi)\right)}{2F_1} & \frac{(B_1 + B_2)\cos(\theta) + F_1}{2F_1} \end{pmatrix}.
\tag{23}
$$

where $F_1$, $B_1$, and $B_2$ defined in Appendix A.

In Fig. 2, the variations of Josephson energies, mutual coupling energy, and temperature on QFI for estimating temperature $T$ at the teleportation destination $\mathcal{F}_T^{out}$ is plotted. This figure demonstrates the high sensitivity of our quantum remote sensing protocol at low temperatures. As can be seen, tuning the Josephson energies at low temperatures improves the sensing sensitivity, while at high temperatures, this sensitivity is suppressed. Furthermore, increasing the mutual coupling energy between qubits improves the remote sensing sensitivity in temperature estimation. It's due to the practical limitations of the structure for two dissimilar coupled qubits under quantum oscillations [62]. In fact, the enhanced sensitivity can be leveraged for applications such as detecting subtle temperature variations in nanoscale devices or monitoring cryogenic processes with high precision. However, the performance of the sensor is also affected by environmental noise and decoherence effects. These factors can limit the achievable sensitivity and introduce uncertainties in temperature estimation. Therefore, mitigating the impact of noise and decoherence is crucial for realizing the full potential of the quantum remote sensing protocol.

Next, we investigate the quantum thermometry scenario in which Alice, possessing a proposed quantum sensor, estimates the temperature directly. In Fig. 3, we illustrate the comparison between the QFI and HSS for estimating temperature as a function of temperature $T$ for the first scenario in which Alice estimates temperature directly by our quantum sensor (Fig. 3(a)) and the second scenario in which Bob estimates temperature remotely by remote estimation without the quantum sensor (Fig. 3(b)). This figure shows that Hilbert-Schmidt speed can also be easily used as a measure to determine the sensitivity of a quantum sensor and, therefore, quantum sensing. Because the qualitative behaviors of both are similar, and the maximum and minimum points of the behaviors coincide. This is due to the many similarities between HSS and QFI of the family of quantum statistical speeds. Refs. [44, 52, 63] study the HSS application to quantum estimation. HSS has also been reported to monitor dynamic non-Markovianity effects [64], global memory effects in noisy channels [65], encrypted information extraction [19, 49], and variations of refractive index [66].

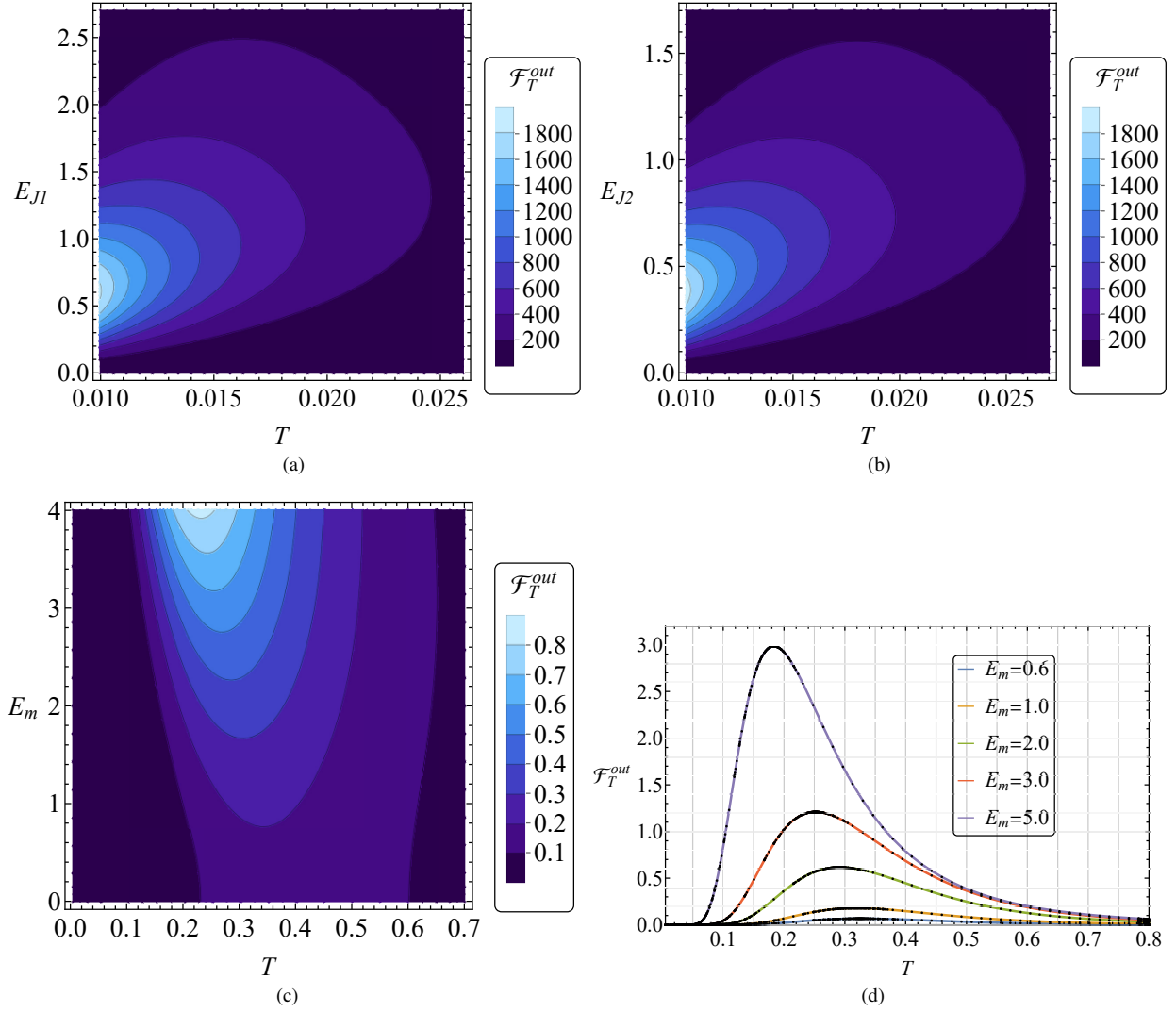

FIG. 2. The QFI for estimating temperature at teleportation destination $\mathcal{F}_T^{out}$ as a function of temperature $T$ for variations of **a)** the Josephson energy of first qubit for $E_{J2} = 0.05, E_m = 4, \theta = \phi = \pi/2$, **b)** the Josephson energy of second qubit for $E_{J1} = 0.06, E_m = 3, \theta = \phi = \pi/2$, **c)** the mutual coupling energy between qubits for $E_{J1} = 2, E_{J2} = 0.8, \theta = \phi/4, \phi = \phi/3$, and **d)** the mutual coupling energy between qubits for $E_{J1} = 1, E_{J2} = 1.3, \theta = \phi = \pi/2$.

Fig. 4 compares the QFI for temperature estimation as a function of temperature $T$ in two scenarios. The QFI demonstrates that directly estimating temperature with the proposed quantum sensor (Scenario 1) is more sensitive than remote temperature estimation without a quantum sensor at the teleportation destination (Scenario 2). This result highlights the advantage of direct quantum sensing for temperature estimation by the quantum sensor. Furthermore, the enhanced sensitivity offered by Scenario 1 translates to a more precise temperature measurement. The QFI quantifies the ultimate precision achievable in estimating a parameter, and a higher QFI value indicates a lower bound on the statistical error. Therefore, the results presented in Fig. 4 imply that the proposed quantum sensor enables temperature measurements with a higher degree of accuracy compared to the remote estimation method. This advantage is particularly relevant in applications where precise temperature monitoring is crucial, such as in nanoscale devices or biological systems.

Figure 5 compares the fidelity of single-qubit quantum thermal teleportation and the HSS for temperature estimation at the teleportation destination across varying temperatures. At low temperatures, the fidelity initially remains stable before decreasing as temperature increases. A comparison of fidelity and HSS reveals that changes in the sign of fidelity correspond to approximately the same points where the sign of HSS changes. Both metrics decrease with increasing temperature. Notably, fidelity exceeds the classical threshold (CT=2/3) and reaches unity at low temperatures, indicating optimal quantum teleportation, but this is suppressed at higher temperatures. Potential causes could include increased thermal noise due to the thermal bath or limitations in the quantum channel's capacity at higher temperatures.

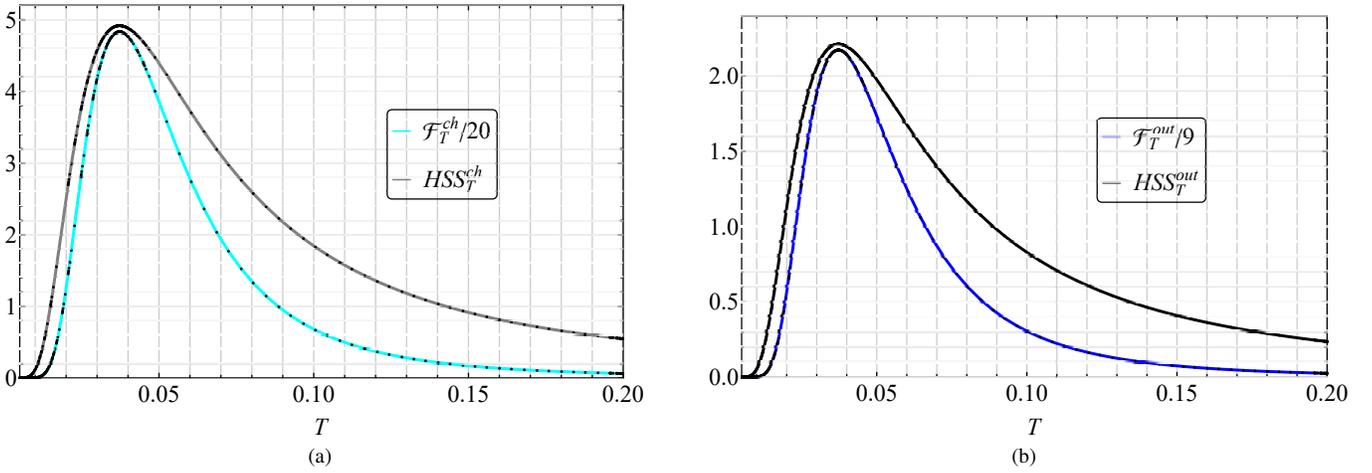

(a)

(b)

FIG. 3. Comparison between QFI and HSS for estimating temperature in **a)** the first scenario when $E_{J1} = 1, E_{J2} = 0.1, E_m = 1$, and **b)** the second scenario $E_{J1} = 1, E_{J2} = 0.1, E_m = 1, \theta = \phi = \pi/2$.

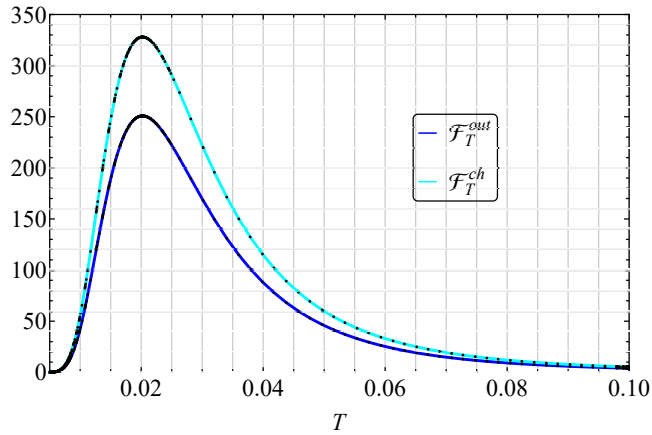

FIG. 4. Comparison QFI for estimating temperature between the first and second scenarios when $E_{J1} = 0.05, E_{J2} = 2, E_m = 1, \theta = \pi/2, \phi = \pi/6$.

## V. CONCLUSIONS

This study presents a comprehensive investigation into quantum thermometry employing state-of-the-art quantum sensor architectures, with an emphasis on the fundamental precision bounds established by quantum estimation theory. The analysis reviews contemporary platforms utilizing charge-disparate (dissimilar-charge) qubits, highlighting their potential applications in emerging fields such as cryogenic quantum processors. The proposed quantum sensor architecture comprises a bipartite system of charge-variant qubits electrostatically coupled via an on-chip capacitive element. These qubits interact with a thermal reservoir at a predefined temperature, with the system's thermal equilibrium described by a Gibbs thermal state derived from the total Hamiltonian. Notably, one of the qubits incorporates a superconducting quantum interference device (SQUID), while the other remains a conventional charge qubit, resulting in quantum oscillatory phenomena driven by charge and flux dynamics. Operationally, the quantum sensor functions as a probe for precise temperature estimation within the framework of quantum thermometry. To quantify the achievable limits of measurement precision, we employ the quantum Fisher information (QFI) and Hilbert-Schmidt speed (HSS), focusing on optimizing the quantum sensor's sensitivity to temperature variations. Parametric analyses reveal the influence of Josephson coupling energies and mutual capacitive interactions on QFI and HSS, enabling the delineation of optimal regimes conducive to enhanced thermal resolution. A comparative analysis contrasting local, direct tem-perature estimation via the quantum sensor—denoted as Alice—with remote estimation facilitated through quantum teleportation protocols—denoted as Bob—indicates that direct sensing generally yields superior sensitivity. This is attributed to the direct interaction between the sensor and the thermal environment, which minimizes noise introduction inherent in teleportation-based measurement schemes. Furthermore, the study demonstrates that increasing Josephson energies diminishes sensor sensitivity, whereas enhanced mutual capacitive coupling acts to improve it. These insights underpin potential applications spanning ultra-precise cryogenic thermometry, in vivo nano-thermometry, on-chip hot-spot profiling, distributed environmental monitoring, and point-of-use thermal validation.

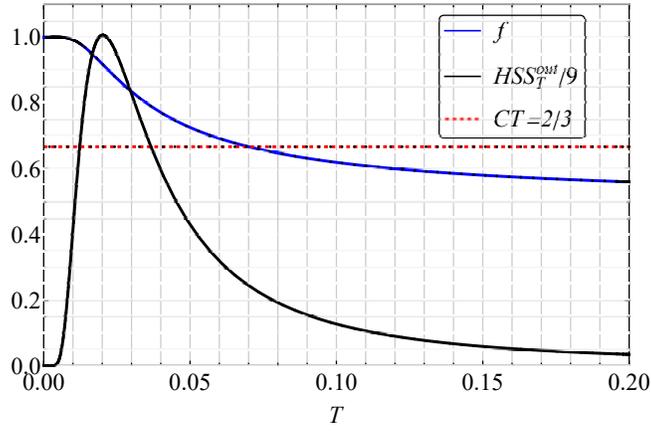

FIG. 5. Comparison between fidelity $f$ and the HSS for estimating temperature $HSS^{out}_T$, is plotted as a function of temperature $T$ when $E_{J1} = 1, E_{J2} = 0.05, E_m = 0.5, \theta = \pi/2, \phi = \pi$. Here, $CT = 2/3$ represents the threshold of the maximum classical fidelity.

## APPENDIX A: EXTRACTION OF THERMAL DENSITY MATRIX

Assuming $n_{g1} = n_{g2} = 0.5$, $E_{00} = E_{11}$ and $E_{01} = E_{10}$, and Considering Eqs. (20-22), we can compute the thermal density matrix as follows (according to what we have obtained in Ref. [62])

$$\rho(T) = \begin{bmatrix} \rho_{11} & \rho_{12} & \rho_{13} & \rho_{14} \\ \rho_{12} & \rho_{22} & \rho_{23} & \rho_{13} \\ \rho_{13} & \rho_{23} & \rho_{22} & \rho_{12} \\ \rho_{14} & \rho_{13} & \rho_{12} & \rho_{11} \end{bmatrix}, \tag{A-1}$$

where the elements of the density matrix are given by

$$\rho_{11} = \rho_{44} = -\frac{D(B_1 + B_2 - F_1)}{2},$$

$$\rho_{22} = \rho_{33} = \frac{D(B_1 + B_2 + F_1)}{2},$$

$$\rho_{14} = \rho_{41} = -\frac{D(-B_1 + B_2 + F_2)}{2},$$

$$\rho_{23} = \rho_{32} = \frac{D(-B_1 + B_2 - F_2)}{2},$$

$$\rho_{12} = \rho_{21} = \rho_{34} = \rho_{43} = D(C_1 + C_2),$$

$$\rho_{13} = \rho_{31} = \rho_{24} = \rho_{42} = D(-C_1 + C_2),$$

$$D = \frac{1}{2F_1}, F_1 = \cosh[A_1] + \cosh[A_2],$$

$$F_2 = \cosh[A_1] - \cosh[A_2], A_1 = \frac{R_1}{4T}, A_2 = \frac{R_2}{4T},$$

$$R_1 = \sqrt{4(E_{J1} - E_{J2})^2 + E_m^2}, R_2 = \sqrt{4(E_{J1} + E_{J2})^2 + E_m^2},$$

$$B_1 = \frac{E_m \sinh(A_1)}{R_1}, B_2 = \frac{E_m \sinh(A_2)}{R_2},$$

$$C_1 = \frac{(E_{J1} - E_{J2})\sinh(A_1)}{R_1}, C_2 = \frac{(E_{J1} + E_{J2})\sinh(A_2)}{R_2}.$$

(A-2)

## ABBREVIATIONS

Nitrogen-vacancy (NV); Possible positive-operator-valued measures (POVMs); Symmetric logarithmic derivative (SLD); Quantum Fisher information (QFI); Hilbert-Schmidt speed (HSS); Cramér-Rao (CR); Superconducting quantum interference device (SQUID);

## AUTHOR CONTRIBUTIONS

Practical research was conducted by S.M.H., A.P.K., S.GH., and M.N. Interpretations and comparison of results and writing of the article were done by S.M.H., M.N., A.P.K, S.GH., A.A., and S.A.K. with the help of J.S.Y. The article was reviewed and edited by J.S.Y.

## COMPETING INTERESTS

The authors declare no competing interests.

## DATA AVAILABILITY STATEMENT

All data generated or analyzed during this study are included in this paper.

## ADDITIONAL INFORMATION

Correspondence and requests for materials should be addressed to S.M.H. and J.S.Y.

## DECLARATIONS

### Ethical Approval and Consent to participate

Not applicable.

### Consent for publication

The Authors confirm: that the work described has not been published before; that it is not under consideration for publication elsewhere; that its publication has been approved by all the authors.

### Funding

Not applicable.